\documentclass[%
 reprint, 
 superscriptaddress, 
 amsmath,amssymb,
 aps, prb, 
]{revtex4-2}
\usepackage{orcidlink}
\usepackage{graphicx}
\usepackage{dcolumn}
\usepackage{bm}
\usepackage[normalem]{ulem}
\usepackage{xcolor}
\usepackage{array,mathtools,amssymb,booktabs}

\newcommand{\tr}[1]{\textcolor{black}{#1}}
\newcommand{\tb}[1]{\textcolor{black}{#1}}

\begin{document}


\title{\textbf{Anomalous temperature dependence of the band gap in TiNCl}
}%

\author{Masaru Oda\,\orcidlink{0000-0002-5667-2856}}
\email{odamasa@mns.kyutech.ac.jp}

\affiliation{Graduate School of Engineering,
Kyushu Institute of Technology, Kitakyushu 804-8550, Japan}

\author{Taisei Ukon}
\author{Yuma Mori}
\author{Kazuma Nakamura\,\orcidlink{0000-0001-8402-4031}}
\author{Masashi Tanaka\,\orcidlink{0000-0002-4323-5999}}

\affiliation{Graduate School of Engineering,
Kyushu Institute of Technology, Kitakyushu 804-8550, Japan}

\author{Hisao Kondo}
\affiliation{Graduate School of Science and Engineering,
Ehime University, Matsuyama 790-8577, Japan}

\author{Aishi Yamamoto}
\affiliation{Department of Mechanical Engineering and Informatics,
Hiroshima Institute of Technology, Hiroshima 731-5193, Japan}

\date{\today}

\begin{abstract}
We report an unusual nonmonotonic temperature dependence of the optical band gap in the layered nitride halide TiNCl, revealed by temperature-dependent photoluminescence (PL) and diffuse reflectance spectroscopy. The band gap, determined from the PL spectra, decreases with increasing temperature at low temperatures, reaches a minimum near 50~K, and then increases at higher temperatures. This behavior contrasts with the monotonic decrease typically observed in conventional semiconductors. The observed temperature dependence is well reproduced by a Bose--Einstein oscillator model with two characteristic phonon energies \textcolor{black}{of approximately 8.9~meV and 1.2~meV}. From temperature-dependent x-ray diffraction measurements and first-principles calculations using the experimental lattice parameters, we find that the contribution of thermal expansion is negligibly small compared with the observed band-gap variation. \textcolor{black}{Furthermore, first-principles phonon calculations for the optimized crystal structure reveal peaks in the phonon density of states near these two characteristic energies.} These results suggest that electron--phonon interactions play an important role in the anomalous temperature dependence of the optical band gap in TiNCl\textcolor{black}{, with low- and higher-energy phonons contributing to the low-temperature decrease and the higher-temperature increase, respectively}.
\end{abstract}

\maketitle


\section{Introduction}
\label{sec:intro}

Layered metal nitride halides
$M\mathrm{N}X$
($M$ = Ti, Zr, Hf, $X$ = Cl, Br, I)
have attracted considerable attention because they exhibit superconductivity upon carrier doping, while their undoped states are nonmetallic~[\citenum{Yamanaka1998}\nocite{Yamanaka2010}--\citenum{Tanaka2022Review}].
The superconducting mechanism has not yet been fully clarified, and extensive studies have been carried out on these materials~\cite{Kasahara2015}.
More recently, first-principles studies of the electronic states, optical properties, and dielectric properties of both monolayer and bulk undoped $M\mathrm{N}X$ compounds have also progressed~\cite{RostamiOsanloo2022,Huang2023,Yun2017,Gomez-Bastidas2025}.
In contrast, although many undoped $M\mathrm{N}X$ compounds possess semiconducting electronic structures~\cite{RostamiOsanloo2022,Huang2023,Yun2017}, experimental optical studies commonly employed in semiconductor research remain limited.
The optical properties of undoped TiNCl also remain largely unexplored, while its electronic states have been investigated by photoemission and scanning tunnelling spectroscopy~\cite{Kataoka2019,Sugimoto2023}.

In general, optical measurements of semiconductors provide access to the band-gap energy, and its temperature dependence offers important information on the electronic structure and electron-phonon interactions in a material~\cite{Cardona2014}.
It is well known that the temperature dependence of the band gap is mainly governed by two factors: electron--phonon interactions and thermal expansion~\cite{Cardona1989,ODonnell1991a}.
Because these contributions can lead to similar temperature dependences~\cite{ODonnell1991a}, it is important to distinguish \textcolor{black}{ their respective roles to understand the physical origin of the band-gap variation}.

In this study, we investigated the temperature dependence of the optical band gap in layered TiNCl by means of temperature-dependent photoluminescence, diffuse reflectance spectroscopy, and temperature-dependent x-ray diffraction. We found that the optical band gap of TiNCl exhibits a nonmonotonic temperature dependence: it decreases at low temperatures, reaches a minimum near 50~K, and then increases at higher temperatures. By comparing the experimental results with first-principles calculations based on the experimentally determined lattice parameters, we evaluated the contribution of thermal expansion \textcolor{black}{and found it to be negligibly small compared with the observed band-gap variation. We further examined the origin of this anomalous temperature dependence by comparing the characteristic phonon energies extracted using a two-oscillator Bose--Einstein model~\cite{Gobel1998, NoteBE} with the phonon density of states (DOS) obtained from first-principles phonon calculations for the optimized crystal structure.}

\tr{\section{Methods}}

\subsection{Experimental Methods} \label{sec:exp}
TiNCl samples were synthesized following the procedure for sample A reported in Ref.~\cite{Tanaka2022} (see also Ref.~\cite{Tanaka2026} for a recent review of TiNCl synthesis). The resulting powder was pressed into pellets using a hydraulic press and used for the optical measurements.

Photoluminescence (PL) spectra were measured using a continuous-wave laser at 405~nm over a temperature range of 6--300~K, with an excitation intensity of 2.4~W/cm$^2$ at the sample. Diffuse reflectance (DR) spectra were measured using a white-light source, and the diffusely scattered light from the sample was detected. The measured reflectivity $R$ was converted into a quantity proportional to the absorption coefficient using the Kubelka--Munk function, $F(R)=(1-R)^2/(2R)$~\cite{Kubelka}. The DR measurements were performed at 6 and 300~K. Temperature-dependent powder x-ray diffraction measurements were carried out using a Rigaku SmartLab diffractometer equipped with a cryostat/furnace system.

\textcolor{black}{}

\tb{\subsection{{\em Ab initio} calculations}\label{sec:DFT}
{\em Ab initio} density functional calculations for the $\alpha$ form of TiNCl were performed with Quantum Espresso package \cite{QE-2009,QE-2017}. We use the Perdew-Burke-Ernzerhof type~\cite{Perdew_1996} for the exchange-correlation functional, and the norm-conserving pseudopotentials with the code ONCVPSP (Optimized Norm-Conserving Vanderbilt PSeudopotential)~\cite{Hamann2013,Kleinman_1982,Troullier_1991} obtained from the PseudoDojo~\cite{Setten_2018,about-ONCV-QE}. The kinetic energy cutoff is set to be 100 Ry for the wave functions and 400 Ry for the charge density. 
The crystal structures were optimized with a total energy convergence better than 1 meV/atom. We use a 10$\times$10$\times$10 Monkhorst–Pack $k$-mesh for this optimization. The phonon modes were calculated using the density-functional linear-response method~\cite{Baroni_1987,Giannozzi_1991}. The calculation conditions are as follows: 16$\times$16$\times$16 k-mesh and 8$\times$8$\times$3 $q$-mesh are used for the electronic structure and phonon mode calculations, respectively. We carefully checked the numerical convergence of the low-energy phonon modes and confirmed that these computational conditions yield stable phonon dispersions without imaginary frequencies.
}

\textcolor{black}{To evaluate the thermal-expansion contribution
to the band gap, electronic band-structure calculations were
performed using the experimental lattice parameters at each
temperature. The internal atomic coordinates were optimized
while keeping these lattice parameters fixed.
The electronic band structure, phonon dispersion, and phonon
DOS were calculated using the optimized
crystal structure.}

\section{Results and Discussion}
\label{sec:res}
{\color{black}
\subsection{Optical properties}
}
\label{sec:opt}
\begin{figure}[tp]
\centering
\includegraphics[width=8.5cm,trim=0cm 12.7cm 12.6cm 0cm,clip]{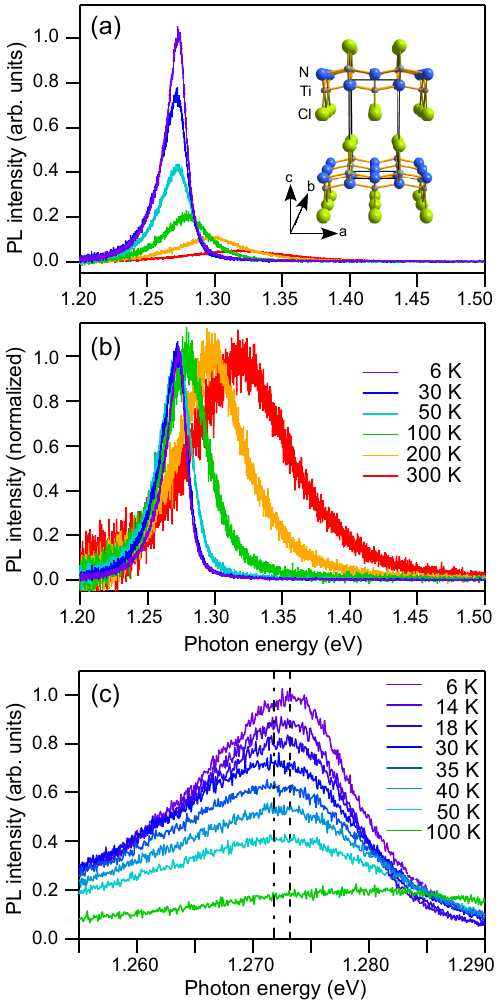}
\caption{
Temperature-dependent photoluminescence (PL) spectra of TiNCl.
(a) Representative PL spectra measured over 6--300~K.
The inset shows the crystal structure of TiNCl with the crystallographic axes indicated.
(b) The same spectra normalized by their peak intensity.
\textcolor{black}{
(c) An expanded view of the low-temperature PL spectra (6--100~K)
from the same data set as in (a).
The dashed and dash-dotted vertical lines indicate the approximate PL peak positions at 6~K and 30--35~K, respectively, and serve as guides to the eye to highlight the initial redshift of the PL peak with increasing temperature.
}}
\label{fig:Fig1}
\end{figure}

Figure~\ref{fig:Fig1}(a) shows representative PL spectra of TiNCl measured over a temperature range of 6--300~K. The inset displays the layered crystal structure of TiNCl  in the $\alpha$ form, the only experimentally known polymorph of this compound~\cite{Tanaka2022Review}. An emission band is observed throughout the measured temperature range. The peak position and spectral shape evolve systematically with increasing temperature. Figure~\ref{fig:Fig1}(b) shows the same spectra normalized by their peak intensity, which more clearly highlights the temperature evolution of the spectral position and spectral shape, particularly on the high-energy side.  \textcolor{black}{At higher temperatures, the PL spectra exhibit a blueshift with increasing temperature [Figs.~\ref{fig:Fig1}(a) and \ref{fig:Fig1}(b)].} The low-temperature PL spectra in the 6--100~K range are shown \textcolor{black}{on an expanded photon-energy scale} in \textcolor{black}{ Fig.~\ref{fig:Fig1}(c)}. \textcolor{black}{In contrast, a small redshift of the PL peak is observed with increasing temperature in the lowest-temperature region. Details of the optical measurement procedures are provided in Sec.~S1 of the Supplemental Material~\cite{SM}.}

To determine the band-gap energy from the PL spectra, we analyzed the spectra using a thermal-distribution model in an interband-transition picture. The observed PL spectrum $P(E)$ was modeled as $P(E)=a\,[I(E)\ast G(E)]$, where $I(E)$ is the ideal spectral distribution reflecting the thermalized carrier distribution before additional spectral broadening. The function $G(E)$ represents additional broadening described by a Gaussian function, and $a$ is an intensity parameter. Assuming that the photoexcited carriers are \textcolor{black}{ in thermal equilibrium with the lattice at temperature $T$} and follow a Boltzmann distribution, the ideal spectrum is written as $I(E)\propto D_{\mathrm{2D}}\exp[-(E-E_{\mathrm g})/(k_{\mathrm B}T)]$, \textcolor{black}{for $E\ge E_{\mathrm g}$, with $I(E)=0$ for $E<E_{\mathrm g}$}, where $D_{\mathrm{2D}}$ is the two-dimensional \textcolor{black}{joint} density of states, assumed to be energy-independent, $E_{\mathrm g}$ is the band-gap energy, and $k_{\mathrm B}$ and $T$ are the Boltzmann constant and absolute temperature, respectively. This form describes the high-energy tail of the PL spectrum arising from thermally distributed carriers. The spectral broadening was modeled by $G(E)\propto \exp(-E^{2}/b^{2})$, where $b$ is a broadening parameter.

The model function $P(E)$ was fitted to the PL spectra at each temperature using a least-squares fitting procedure. In the fitting, $E_{\mathrm g}$, $a$, and $b$ were treated as adjustable parameters; the temperature dependence of $E_{\mathrm g}$ was extracted from the fitting results. The \textcolor{black}{ excitation density used in these measurements lies within the} weak excitation \textcolor{black}{ regime}, where the PL intensity \textcolor{black}{ scales} approximately linearly with excitation density \textcolor{black}{and the spectral shape remains unchanged}; details are provided in Sec.~S\textcolor{black}{2}.

\begin{figure}[t]
\centering
\includegraphics[width=8.5cm,trim=0cm 16.0cm 12.6cm 0cm,clip]{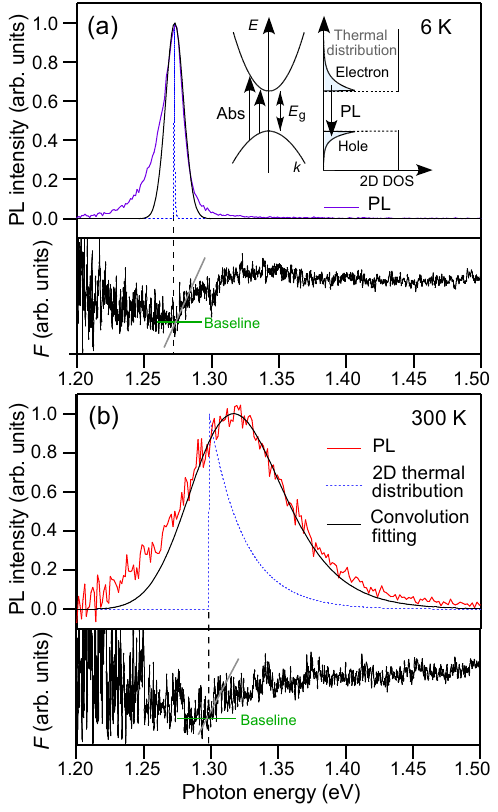}
\caption{
\textcolor{black}{}
\textcolor{black}{PL and diffuse reflectance spectra of TiNCl at (a) 6~K and (b) 300~K.}
\textcolor{black}{}
\textcolor{black}{}
\textcolor{black}{Upper panels: Measured PL spectra (purple in (a) and red in (b)) analyzed using a thermal-distribution model, with total fits shown as black solid curves and thermal-distribution components as blue dotted curves.}
Vertical dashed lines \textcolor{black}{in the upper panels} mark the band-gap energies \textcolor{black}{extracted from the PL fits}.
Lower panels: corresponding $F(R)$ spectra from diffuse reflectance measurements.
Inset in (a): schematic illustration of the \textcolor{black}{PL} model.
}
\label{fig:Fig2}
\end{figure}

Fitting results for the PL spectra at 6 and 300~K are shown in Figs.~\ref{fig:Fig2}(a) and \ref{fig:Fig2}(b), respectively. In both cases, the measured PL spectra are well reproduced by the convolution model over the fitting range. In the fitting, particular emphasis was placed on the high-energy side of the PL spectrum, which more directly reflects the thermal carrier distribution, while the low-energy side may contain additional emission components. The inset shows the model used for the analysis, based on thermally distributed carriers in a two-dimensional \textcolor{black}{joint} density of states. The vertical dashed lines indicate the band-gap energies determined from the PL fitting. The lower panels show the corresponding $F(R)$ spectra obtained from DR measurements, which provide complementary information on the absorption edge. The correspondence between the PL-derived band-gap energies and the absorption-edge features in the $F(R)$ spectra indicates that the PL emission can be understood in terms of an interband-transition picture. At 6~K, the PL spectrum is relatively sharp and the extracted band-gap energy is located close to the emission peak. At 300~K, the PL spectrum becomes broader and exhibits a more pronounced thermal tail, which is mainly governed by the thermal-distribution component of the model together with Gaussian broadening. Additional fitting results at selected temperatures\textcolor{black}{} are provided in Sec.~S\textcolor{black}{3}.

\begin{figure}[t]
\centering
\includegraphics[width=8.5cm,trim=0cm 18.9cm 12.5cm 0cm,clip]{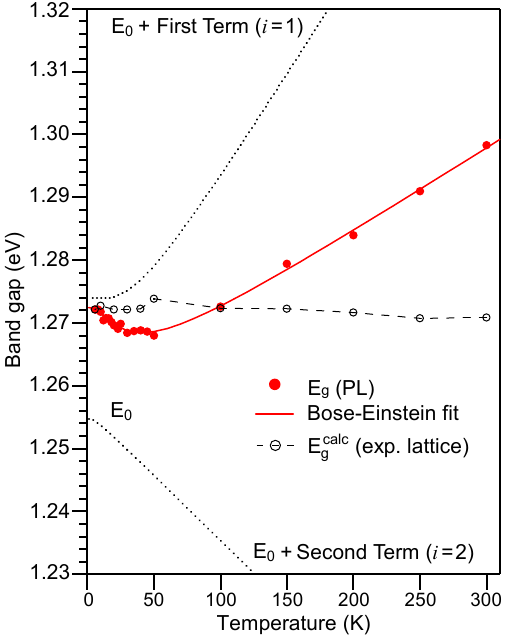}
\caption{
Temperature dependence of the band-gap energy $E_\mathrm{g}$ extracted from the PL analysis. Closed circles show the experimental values obtained from the PL fitting. The solid line represents a fit using a two-oscillator Bose--Einstein model. The dashed line shows the calculated band gap $E_\mathrm{g}^{\mathrm{calc}}$ estimated from first-principles calculations using the measured lattice parameters, after applying a rigid offset to match the experimental value at low temperature. The two dotted curves represent $E_0$ plus the first and second terms in Eq.\textcolor{black}{} \textcolor{black}{(2)}, respectively.
}
\label{fig:Fig3}
\end{figure}

\begin{figure}[t]
\centering
\includegraphics[width=8.5cm,trim=0cm 18.6cm 12.6cm 0cm,clip]{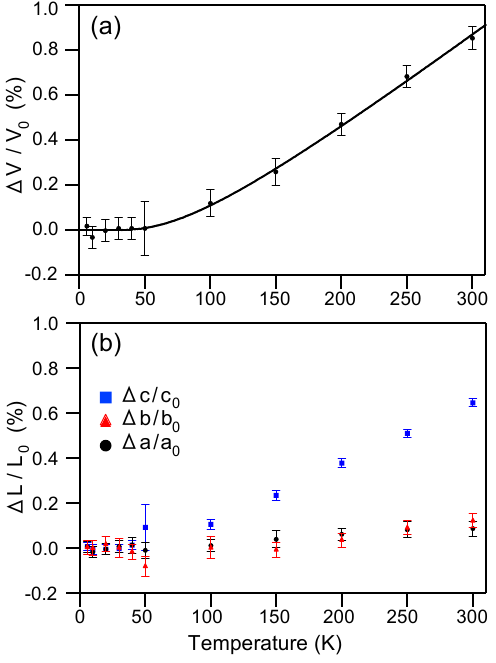}
\caption{
Temperature dependence of the lattice parameters of TiNCl determined from powder x-ray diffraction measurements.
(a) Relative change in the unit-cell volume, \textcolor{black}{$\Delta V/V_0$}.
The solid line is a fit using a one-oscillator Bose--Einstein-type model. 
(b) Relative changes in the lattice parameters, $\Delta a/a_0$, $\Delta b/b_0$, and $\Delta c/c_0$. \textcolor{black}{Here, $a_0$, $b_0$, and $c_0$ are defined as the respective averages of the lattice parameters $a$, $b$, and $c$ below 40~K, where all three lattice parameters can be regarded as constant within experimental uncertainty.}
}
\label{fig:Fig4}
\end{figure}

{\color{black}
\subsection{Temperature Dependence of the Band Gap}
}
\label{sec:bnd}

Figure~\ref{fig:Fig3} summarizes the temperature dependence of the optical band-gap energy in TiNCl extracted from the PL analysis. The closed circles represent the band-gap energies at each temperature obtained from the fitting of the PL spectra. As the temperature increases, $E_{\mathrm g}$ initially decreases, reaches a minimum around 50~K, and then increases at higher temperatures. This nonmonotonic behavior is in clear contrast to the monotonic decrease typically observed in conventional semiconductors~\cite{Cardona2014}.

To examine whether this behavior can be explained by lattice thermal expansion, we compared the experimental results with first-principles calculations using the measured lattice parameters at each temperature. The measured lattice parameters are summarized in Fig.~\ref{fig:Fig4}, which shows the relative change in the unit-cell volume and the relative changes in the lattice parameters. \textcolor{black}{Here, $V_0$ is an effective 0~K reference volume, defined as the average volume below 40~K, where the volume can be regarded as constant within experimental uncertainty. The experimental volume change is defined as $\Delta V=V-V_0$, where $V$ is the volume at temperature $T$.} 
\textcolor{black}{The Bose--Einstein-type fit shown by the black line
in Fig.~\ref{fig:Fig4}(a) is described in the Appendix.}
The temperature-dependent x-ray diffraction patterns and the refined lattice parameters are provided in Sec.~S\textcolor{black}{4}. The largest change is observed along the $c$ axis, whereas the changes along the $a$ and $b$ axes are much smaller. This anisotropic thermal expansion was incorporated into the calculations. Because density functional theory (DFT) underestimates the absolute band gap, we consider only the temperature-dependent change obtained from the calculations and align it with the experimental value at low temperature by applying a rigid offset\textcolor{black}{,} \begingroup
\color{black}
as follows:
\begin{equation}
E_{\mathrm{g}}^{\mathrm{calc}}(T)
=
\left[
E_{\mathrm{g}}^{\mathrm{DFT}}(T)
-
E_{\mathrm{g}}^{\mathrm{DFT}}(5.7\,\mathrm{K})
\right]
+
E_{\mathrm{g}}^{\mathrm{exp}}(6\,\mathrm{K}),
\label{eq:Eg_calc}
\end{equation}
where $E_{\mathrm{g}}^{\mathrm{DFT}}(T)$ is the band gap calculated
using the experimental lattice parameters at temperature $T$,
and $E_{\mathrm{g}}^{\mathrm{exp}}(6\,\mathrm{K})$ is the experimental
band gap obtained from the PL analysis at 6~K.
The lowest temperatures of the x-ray diffraction and optical
measurements are 5.7 and 6~K, respectively; this small temperature
difference is assumed to have a negligible effect on the comparison.
\endgroup
The resulting band gap, denoted as $E_\mathrm{g}^{\mathrm{calc}}$, is shown by the dashed line in Fig.~\ref{fig:Fig3}. \textcolor{black}{} The calculated contribution of thermal expansion is negligibly small and does not reproduce the observed nonmonotonic temperature dependence, indicating that thermal expansion is not the main origin of the observed behavior.

To examine the contribution of electron--phonon interactions to the observed band-gap variation, we employ a semi-empirical two-oscillator Bose--Einstein model~[\citenum{Gobel1998}, \citenum{NoteBE},
\textcolor{black}{\citenum{Bhosale2012}}],
\begin{equation}
E_{\mathrm g}(T)=E_0+\sum_{i=1}^{2} A_i
\left[
\frac{1}{\exp(\textcolor{black}{E_i}/k_{\mathrm B}T)-1}+\frac{1}{2}
\right].
\label{eq:BE_model}
\end{equation}
Here, $E_0$ is the band-gap energy in the limit of zero-amplitude atomic motion, $A_i$ denote the signed energy contributions of each oscillator to the band-gap renormalization, and $\textcolor{black}{E_i}$ are the effective phonon energies. The complex contributions of all phonon modes are approximated by two effective phonon modes interacting with the electronic states.

The experimental data are well reproduced by this model, as shown by the solid line in Fig.~\ref{fig:Fig3}.
The fitting yields $A_1 = 35.6$~meV and $A_2 = -3.0$~meV, together with two effective phonon energies, $E_1 = 8.9$~meV and $E_2 = 1.2$~meV; these parameters and the corresponding first-principles results described below are summarized in Table~\ref{model_param}.
The dotted curves in Fig.~\ref{fig:Fig3} represent the individual contributions of the $i=1$ and $i=2$ oscillators, respectively.
The term in brackets in Eq.~(\ref{eq:BE_model}) corresponds to the phonon occupation number including the zero-point vibration. At very low temperatures, this function increases only gradually with temperature. As the temperature approaches the characteristic energy scale set by each effective phonon energy $\textcolor{black}{E_i}$, the increase becomes more pronounced, and at higher temperatures it approaches an almost linear dependence on temperature. As seen from the two dotted curves in Fig.~\ref{fig:Fig3}, the first oscillator ($i=1$) exhibits a noticeable increase in its contribution in the temperature range of approximately 20--70~K. In contrast, because of the much smaller effective phonon energy, the second oscillator ($i=2$) reaches this transition regime at very low temperatures and contributes nearly linearly over a wide temperature range. Accordingly, $\textcolor{black}{E_2}$ should be regarded as an effective low-energy scale rather than a precisely determined phonon energy.
Since the bracketed function in Eq.~(\ref{eq:BE_model}) is a monotonically increasing function of temperature, a single-term fit cannot reproduce the observed nonmonotonic behavior. This indicates that the nonmonotonic behavior arises from the competition between two contributions with opposite signs, namely those associated with $A_1$ and $A_2$.
Similar competing contributions leading to anomalous band-gap shifts
have been reported in certain materials such as CuCl~\cite{Gobel1998} and chalcopyrite semiconductors~\cite{Bhosale2012}.

\begin{table}[htpb]
\caption{\tb{Parameters of a semi-empirical two-oscillator Bose--Einstein (BE) model in Eq.~(\ref{eq:BE_model}). $E_0$ is given in eV, and $E_1$ and $E_2$ are given in meV. $A_1$ and $A_2$ are given in meV. For the {\em ab initio} calculations, $E_0$ denotes the calculated band gap, while $E_1$ and $E_2$ denote the characteristic energies in the phonon density of states shown in Figs.~\ref{fig_phonon}(c) and \ref{fig_phonon_zoom}(b), respectively.}}
\begin{center}
\begin{tabular}{l@{\ \ \ }c@{\ \ \ }c@{\ \ \ }c@{\ \ \ }c@{\ \ \ }c}
\hline \hline \\ [-5pt]
model
& \textcolor{black}{$E_0$}
& \textcolor{black}{$A_1$}
& $A_2$
& $E_1$
& $E_2$ \\ \hline \\ [-5pt]
\tb{BE} [Eq.~(\ref{eq:BE_model})]
& \tb{1.256} 
& 35.6
& \textcolor{black}{$-2.98$}
& \textcolor{black}{8.93}
& \textcolor{black}{1.24} \\ [+5pt]
\tb{{\em Ab initio}}
& \tb{0.623}
& -
& -
& \tb{10.6}
& \tb{1.24} \\
[+5pt] \hline \hline
\end{tabular}
\end{center}
\label{model_param}
\end{table}

\textcolor{black}{To provide a basis for discussing the origin of the anomalous band-gap shift and interpreting the characteristic phonon energies extracted from the two-oscillator model, we next present the calculated electronic band structure and phonon properties of TiNCl.}

\tb{\subsection{{\em Ab initio} band and phonon structures}}\label{sec:band}

\begin{figure}[b]
\centering
\includegraphics[width=\linewidth]{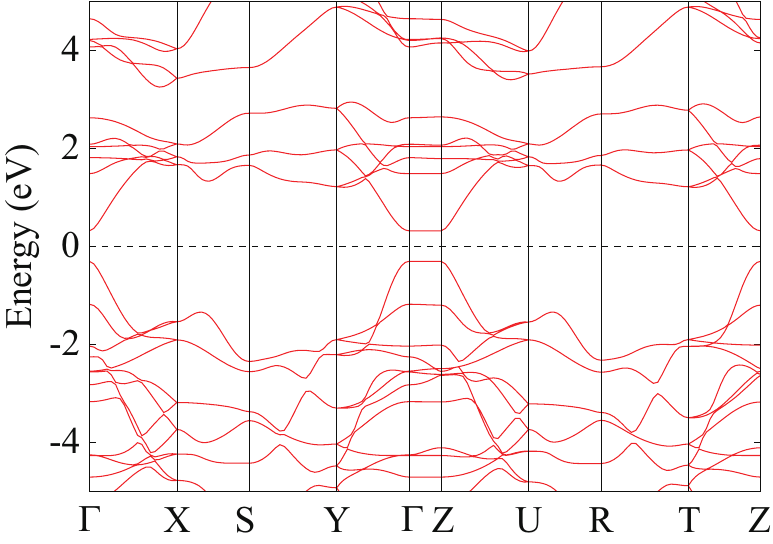}
\caption{
\tb{{\em Ab initio} density functional band structure, where $\Gamma$= (0, 0, 0), X = (1/2, 0, 0), S = (1/2, 1/2, 0), Y = (0, 1/2, 0), Z = (0, 0, 1/2), U = (1/2, 0, 1/2), R = (1/2, 1/2, 1/2), and T = (0, 1/2, 1/2), and the coordinates are represented in terms of basic vectors of the reciprocal lattice of orthorhombic lattice.}
}
\label{Fig_band}
\end{figure}
\begin{figure*}[htpb]
\centering
\includegraphics[width=\linewidth]{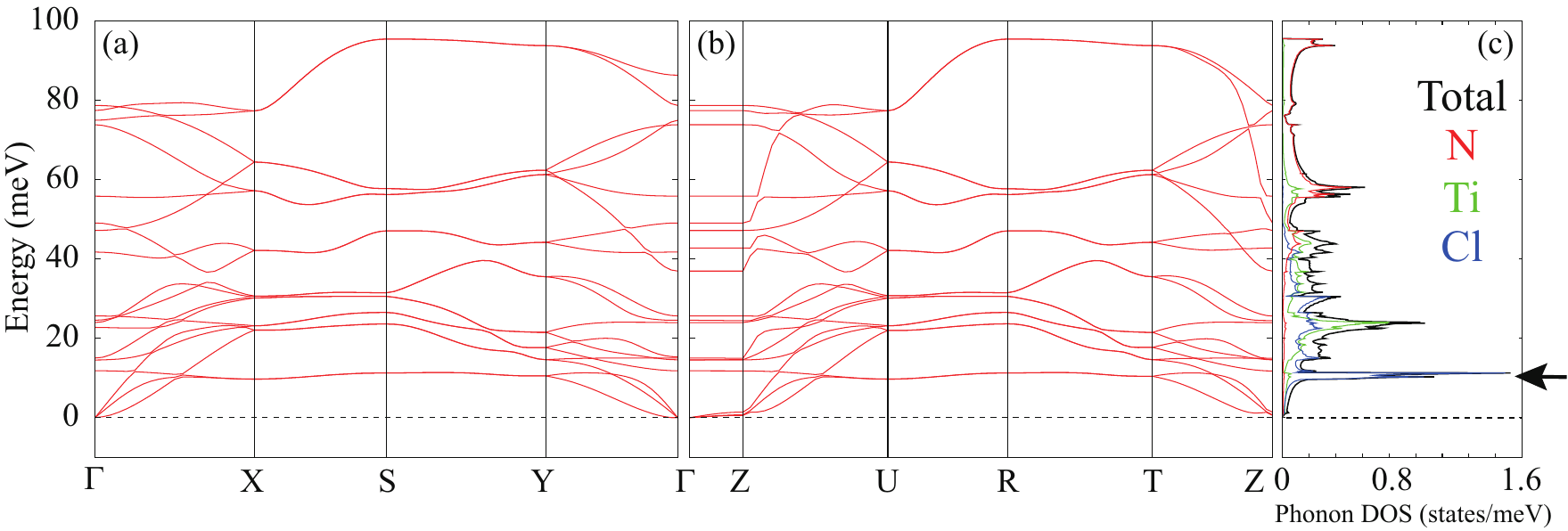}
\caption{
\tb{Calculated phonon dispersion along (a) $\Gamma \to {\rm X} \to {\rm S} \to {\rm Y} \to \Gamma$ and (b) $\Gamma \to {\rm Z} \to {\rm U} \to {\rm R} \to {\rm T} \to {\rm Z}$, where $\Gamma$= (0, 0, 0), X = (1/2, 0, 0), S = (1/2, 1/2, 0), Y = (0, 1/2, 0), Z = (0, 0, 1/2), U = (1/2, 0, 1/2), R = (1/2, 1/2, 1/2), and T = (0, 1/2, 1/2), and the coordinates are represented in terms of basic vectors of the reciprocal lattice of orthorhombic lattice. (c) Calculated phonon density of states (DOS). The arrow indicates the minimum between the two distinct peaks around 10 meV. 
}
}
\label{fig_phonon}
\end{figure*}
\tb{Figure~\ref{Fig_band} shows the calculated electronic band structure. The calculated band gap is 0.623 eV, which is smaller than the experimental value of 1.27 eV. This underestimation is attributed to the use of the PBE functional. The electronic structure exhibits a quasi-two-dimensional character, as indicated by the weak band dispersion along the $\Gamma$--Z direction. The overall band structure is consistent with that obtained in a previous study~\cite{Quan_2011}, except for the magnitude of the band gap.
}

\tb{Figure~\ref{fig_phonon} shows the calculated phonon dispersions and density of states. Panel (a) shows the in-plane phonon dispersion in the $q_z=0$ plane. Panel (b) shows the out-of-plane dispersion along the $\Gamma$--Z direction, followed by the in-plane dispersion in the $q_z=1/2$ plane. Panel (c) shows the total and atom-projected phonon density of states (DOS). The arrow indicates the minimum between the two distinct peaks around 10 meV.}
\begin{figure}[h]
\centering
\includegraphics[width=0.88\linewidth]{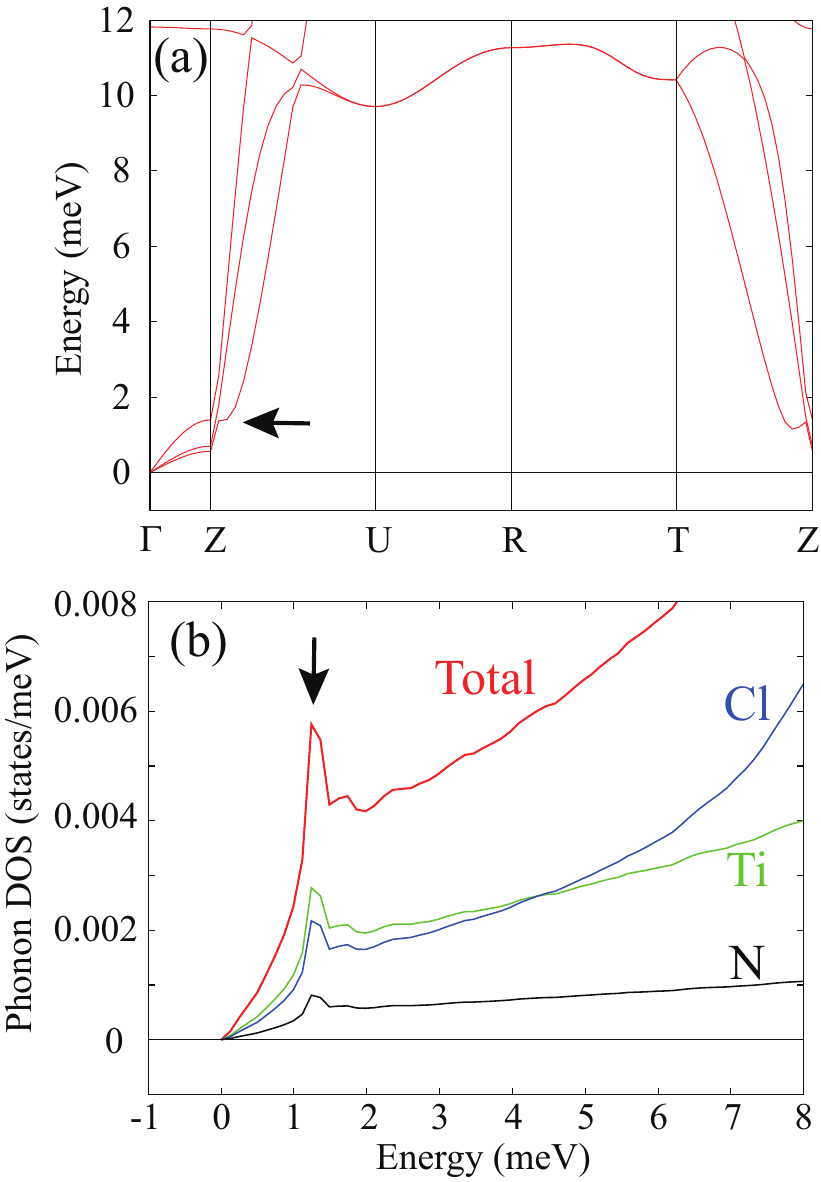}
\caption{
\tb{{\em Ab initio} calculations of the low-energy phonon properties. (a) Phonon dispersion. (b) Total and atom-projected phonon DOS. The dispersion path is the same as that in Fig.~\ref{fig_phonon}. Weakly dispersive acoustic phonon branches are observed along the $\Gamma$--Z direction in the energy range of 0--2 meV. In particular, a nearly flat dispersion appears around 1.2~meV near the Z point, as indicated by the arrow in panel (a). The arrow in panel (b) indicates the corresponding peak in the phonon DOS around 1.2~meV.
}
}
\label{fig_phonon_zoom}
\end{figure}

\tb{Figure~\ref{fig_phonon_zoom} shows enlarged views of the low-energy phonon dispersion and DOS. In panel (a), the phonon dispersion along the out-of-plane direction is shown along the $\Gamma$--Z line. Low-energy phonon branches below approximately 2 meV exhibit only weak dispersion along this direction. The Z--U--R--T--Z path represents the in-plane phonon dispersion in the $q_z=1/2$ plane, where the phonon branches are generally more dispersive. A weakly dispersive branch is also observed around 10 meV. In particular, a nearly flat region appears around the Z point at approximately 1 meV, as indicated by the arrow. This flat dispersion gives rise to an enhanced phonon DOS around this energy. } 

\tb{Panel (b) shows an enlarged view of the low-energy phonon DOS, where a pronounced peak appears around 1.2 meV, as indicated by the arrow. The atom-projected DOS shows that Ti and Cl mainly contribute to the low-energy phonon states below 1.5 meV, while the contribution from N is relatively small. Below the peak, the phonon DOS shows a strongly energy-dependent increase toward the peak. Above approximately 1.5 meV, the DOS increases again with energy, but with a different slope.} 

\tb{To summarize, we identify two characteristic phonon energy scales associated with van Hove-like features in the phonon spectrum. One is the energy scale around 10 meV, which is mainly associated with Cl vibrations, as shown in Fig.~\ref{fig_phonon}(c). The other is the low-energy scale around 1 meV as shown in Fig.~\ref{fig_phonon_zoom}(b), which originates from the weakly dispersive phonon branch around the Z point and gives rise to the pronounced peak in the low-energy phonon DOS.} 


{\color{black}
\subsection{Origin of the Nonmonotonic Band-Gap Shift}
}
\label{sec:origin}
\begingroup
\color{black}
We combine the results of Sec.~III~B with the phonon calculations presented in Sec.~III~C. Since the contribution of lattice thermal expansion is negligibly small, as shown in Sec.~III~B, we focus here on electron--phonon interactions.

The higher effective phonon energy $E_1 = 8.9$~meV obtained from the two-oscillator Bose--Einstein model lies near a band of phonon modes with a double-peak structure in the DOS, marked by the arrow at approximately 10.6~meV in Fig.~6(c).
The atom-projected phonon DOS shown in Fig.~6(c) indicates that this double-peak structure is dominated by Cl vibrations.
Thus, we interpret the anomalous high-energy shift of $E_{\mathrm g}$ above 50~K as arising from the combined contribution of these Cl-dominated phonon modes over a finite energy range, effectively represented by the higher-energy oscillator ($i=1$).

The lower effective phonon energy $E_2 = 1.2$~meV obtained from the same model is close to the DOS peak at 1.2~meV indicated by the arrow in Fig.~7(b), which shows an enlarged view of the low-energy region of the DOS in Fig.~6(c).
The approximately quadratic increase in the DOS at the
lowest energies suggests three-dimensional acoustic phonon
behavior reflecting dispersion in both the in-plane and
interlayer directions.
The approximately linear increase at higher energies
within this low-energy region suggests quasi-two-dimensional
behavior dominated by in-plane dispersion.
We therefore interpret the lower-energy oscillator ($i=2$) as an effective representation of the contribution of phonons around the DOS peak indicated by the arrow in Fig.~7(b) and in the surrounding crossover region associated with the layered structure.

Within the energy range encompassing the two phonon groups discussed above, no other prominent features are apparent in the calculated DOS. Taken together, these results support the interpretation that the low- and higher-energy phonon groups represented by the two effective oscillators make competing electron--phonon contributions with opposite signs, giving rise to the nonmonotonic temperature dependence of the band gap in TiNCl.

Microscopically, the band-gap renormalization induced by
electron--phonon interactions can be understood in terms of the
Fan and Debye--Waller contributions to the electron--phonon
self-energy~\cite{Allen1981,Cardona1989}.
A quantitative evaluation of these contributions to the
band-gap shift remains a subject for future work.
\endgroup

\tb{
Finally, we comment on the thermal expansion effect on the band gap. Thermal expansion is known to contribute to the temperature dependence of the band gap in conventional semiconductors such as GaAs~\cite{ODonnell1991a}. With increasing temperature, the lattice volume expands, which can lead to a reduction of the band gap. Therefore, the effect of thermal expansion should be considered when discussing the temperature dependence of the band gap. We therefore examined whether the experimentally observed temperature dependence of the lattice parameters can account for the band-gap variation in TiNCl. TiNCl exhibits relatively small thermal expansion, as shown in Fig.~\ref{fig:Fig4}. However, the band gap calculated using the experimental lattice parameters at each temperature hardly changes with temperature. This result indicates that the experimentally observed temperature dependence of the band gap cannot be explained by thermal expansion alone.
} 

\section{Conclusion}
\label{sec:con}
In summary, we investigated the temperature dependence of the optical band gap in layered TiNCl by means of temperature-dependent photoluminescence, diffuse reflectance spectroscopy, x-ray diffraction, and first-principles calculations. The band gap extracted from the PL spectra exhibits an unusual nonmonotonic temperature dependence: it decreases at low temperatures, reaches a minimum near 50~K, and increases at higher temperatures. The observed temperature dependence is well reproduced empirically by a \textcolor{black}{two-oscillator} Bose--Einstein \textcolor{black}{} model\textcolor{black}{}. The contribution of thermal expansion, evaluated from x-ray diffraction data and first-principles calculations based on the experimental lattice parameters, is negligibly small. \textcolor{black}{Comparison with the calculated phonon DOS suggests that
the higher effective phonon energy is associated with
Cl-dominated vibrations, whereas the lower one reflects
low-energy acoustic phonons in the crossover region
between three-dimensional and quasi-two-dimensional behavior.
Together, the experimental results and the two-oscillator
analysis support the interpretation that competing
electron--phonon contributions with opposite signs give rise
to the anomalous temperature dependence of the band gap
in TiNCl.
}


\begin{acknowledgments}
The authors thank Y. Matsushita (NIMS) for support with the low-temperature x-ray diffraction measurements. We are grateful to H. Kobayashi \textcolor{black}{(formerly at Kyutech)} for his collaboration in the early stages of this work.\author{Haruya Kobayashi}

\end{acknowledgments}

\appendix*

\begingroup
\color{black}

\section{Bose--Einstein analysis of lattice expansion}
\label{app:lattice}

\subsection{Fitting model and results}
\label{app:lattice_fit}

We examine the temperature dependence of the lattice expansion.
We fitted the data in Fig.~\ref{fig:Fig4}(a)
using a one-oscillator Bose--Einstein-type model:
\begin{equation}
\frac{\Delta V}{V_0}
=
C_0+C_1
\left[
\frac{1}{\exp(E_{\mathrm{eff}}/k_{\mathrm B}T)-1}
+\frac{1}{2}
\right],
\label{eq:BE_lattice_app}
\end{equation}
Here, $E_{\mathrm{eff}}$ is the effective phonon energy
and $C_0$ and $C_1$ are dimensionless coefficients.
In the limit $T\to0$, the bracketed term in
Eq.~(\ref{eq:BE_lattice_app}) approaches $1/2$,
giving the constraint $C_0=-C_1/2$.
Thus, $C_1$ and $E_{\mathrm{eff}}$ are treated as independent
fitting parameters.
As shown by the solid line in Fig.~\ref{fig:Fig4}(a),
the model provides a good fit to the observed temperature
dependence of $\Delta V/V_0$, yielding
$E_{\mathrm{eff}}=20.1$~meV and $C_1=1.024\%$,
with $C_0=-0.512\%$ determined by the constraint.

\subsection{Physical basis and discussion}
\label{app:lattice_basis}

To clarify the physical basis of Eq.~(\ref{eq:BE_lattice_app}),
we consider the Gr\"uneisen approximation~\cite{ODonnell1991a,Allen2020}:
\begin{equation}
B\,\frac{\Delta V^{\mathrm{stat}}}{V_0^{\mathrm{stat}}}
=
\gamma\frac{\overline{E}}{V}.
\label{eq:Gruneisen_app}
\end{equation}
Here, $V_0^{\mathrm{stat}}$ denotes the equilibrium volume
of the static lattice, with both thermal and zero-point
vibrations excluded, and
$\Delta V^{\mathrm{stat}}=V-V_0^{\mathrm{stat}}$. The actual volume at 0~K includes the effects
of zero-point vibrations.
$B$ is the bulk modulus, $\gamma$ is the Gr\"uneisen parameter,
and $\overline{E}$ denotes the mean vibrational energy
of phonons, including the zero-point contribution.
$B$ and $\gamma$ are treated as effective,
temperature-independent parameters reflecting
the anisotropic elastic and vibrational properties.
For small volume changes, $V\simeq V_0^{\mathrm{stat}}$,
so that $\Delta V^{\mathrm{stat}}/V_0^{\mathrm{stat}}$
is approximately proportional to $\overline{E}$.

When $\overline{E}$ is described using a Bose--Einstein-type
model with an effective phonon energy
$E_{\mathrm{eff}}$~\cite{ODonnell1991a},
the relative volume change is given by
\begin{equation}
\frac{\Delta V^{\mathrm{stat}}}{V_0^{\mathrm{stat}}}
=
C_1^{\mathrm{stat}}
\left[
\frac{1}{\exp(E_{\mathrm{eff}}/k_{\mathrm B}T)-1}
+\frac{1}{2}
\right],
\label{eq:BE_static_app}
\end{equation}
where $C_1^{\mathrm{stat}}$ is a dimensionless coefficient.
In the present analysis, we approximate the actual volume
at 0~K by $V_0$, defined as the average volume below 40~K.
Under this approximation, the parameters in Eq.~(\ref{eq:BE_lattice_app})
can be expressed as
$C_0=(V_0^{\mathrm{stat}}-V_0)/V_0$ and
$C_1=(V_0^{\mathrm{stat}}/V_0)C_1^{\mathrm{stat}}$,
both of which are temperature-independent.
This provides a physical basis for applying
the Bose--Einstein-type model to the experimentally
observed lattice expansion.

The fitted $E_{\mathrm{eff}}$ differs from $E_1$ and $E_2$
obtained from the band-gap analysis.
The two models describe different physical quantities: temperature-dependent changes in the equilibrium atomic positions and electron--phonon-induced
band-gap renormalization, although both represent the
contributions of many phonon modes using a small number
of effective oscillators.
Thus, this difference is consistent with the interpretation
of the band-gap variation in terms of electron--phonon
interactions.

\par
\endgroup

\bibliography{TINCL_Bandgap}

\end{document}